\documentstyle[twoside,epsfig]{article}
\catcode`\@=11
\long\def\@makefntext#1{
\protect\noindent \hbox to 3.2pt {\hskip-.9pt  
$^{{\eightrm\@thefnmark}}$\hfil}#1\hfill}		%CAN BE USED 

\def\@makefnmark{\hbox to 0pt{$^{\@thefnmark}$\hss}}	%ORIGINAL 
	
\def\ps@myheadings{\let\@mkboth\@gobbletwo
\def\@oddhead{\hbox{}
\rightmark\hfil\eightrm\thepage}   
\def\@oddfoot{}\def\@evenhead{\eightrm\thepage\hfil
\leftmark\hbox{}}\def\@evenfoot{}
\def\sectionmark##1{}\def\subsectionmark##1{}}

\oddsidemargin=\evensidemargin
\newcounter{sectionc}\newcounter{subsectionc}\newcounter{subsubsectionc}
\renewcommand{\section}[1] {\vspace{12pt}\addtocounter{sectionc}{1} 
\setcounter{subsectionc}{0}\setcounter{subsubsectionc}{0}\noindent 
	{\tenbf\thesectionc. #1}\par\vspace{5pt}}
\renewcommand{\subsection}[1] {\vspace{12pt}\addtocounter{subsectionc}{1} 
	\setcounter{subsubsectionc}{0}\noindent 
	{\bf\thesectionc.\thesubsectionc. {\kern1pt \bfit #1}}\par\vspace{5pt}}
\renewcommand{\subsubsection}[1] {\vspace{12pt}\addtocounter{subsubsectionc}{1}
	\noindent{\tenrm\thesectionc.\thesubsectionc.\thesubsubsectionc.
	{\kern1pt \tenit #1}}\par\vspace{5pt}}
\newcommand{\nonumsection}[1] {\vspace{12pt}\noindent{\tenbf #1}
	\par\vspace{5pt}}

\newcounter{appendixc}
\newcounter{subappendixc}[appendixc]
\newcounter{subsubappendixc}[subappendixc]
\renewcommand{\thesubappendixc}{\Alph{appendixc}.\arabic{subappendixc}}
\renewcommand{\thesubsubappendixc}
	{\Alph{appendixc}.\arabic{subappendixc}.\arabic{subsubappendixc}}

\renewcommand{\appendix}[1] {\vspace{12pt}
        \refstepcounter{appendixc}
        \setcounter{figure}{0}
        \setcounter{table}{0}
        \setcounter{lemma}{0}
        \setcounter{theorem}{0}
        \setcounter{corollary}{0}
        \setcounter{definition}{0}
        \setcounter{equation}{0}
        \renewcommand{\thefigure}{\Alph{appendixc}.\arabic{figure}}
        \renewcommand{\thetable}{\Alph{appendixc}.\arabic{table}}
        \renewcommand{\theappendixc}{\Alph{appendixc}}
        \renewcommand{\thelemma}{\Alph{appendixc}.\arabic{lemma}}
        \renewcommand{\thetheorem}{\Alph{appendixc}.\arabic{theorem}}
        \renewcommand{\thedefinition}{\Alph{appendixc}.\arabic{definition}}
        \renewcommand{\thecorollary}{\Alph{appendixc}.\arabic{corollary}}
        \renewcommand{\theequation}{\Alph{appendixc}.\arabic{equation}}
        \noindent{\tenbf Appendix \theappendixc #1}\par\vspace{5pt}}
\newcommand{\subappendix}[1] {\vspace{12pt}
        \refstepcounter{subappendixc}
        \noindent{\bf Appendix \thesubappendixc. {\kern1pt \bfit #1}}
	\par\vspace{5pt}}
\newcommand{\subsubappendix}[1] {\vspace{12pt}
        \refstepcounter{subsubappendixc}
        \noindent{\rm Appendix \thesubsubappendixc. {\kern1pt \tenit #1}}
	\par\vspace{5pt}}

\newcommand{\textlineskip}{\baselineskip=13pt}
\newcommand{\smalllineskip}{\baselineskip=10pt}

\def\eightcirc{
\begin{picture}(0,0)
\put(4.4,1.8){\circle{6.5}}
\end{picture}}
\def\eightcopyright{\eightcirc\kern2.7pt\hbox{\eightrm c}} 

\newcommand{\copyrightheading}[1]
	{\vspace*{-2.5cm}\smalllineskip{\flushleft
	{\footnotesize International Journal of Modern Physics A, #1}\\
	{\footnotesize $\eightcopyright$\, World Scientific Publishing
	 Company}\\
	 }}

\def\abstracts#1#2#3{{
	\centering{\begin{minipage}{4.5in}\baselineskip=10pt\footnotesize
	\parindent=0pt #1\par 
	\parindent=15pt #2\par
	\parindent=15pt #3
	\end{minipage}}\par}} 

\renewenvironment{thebibliography}[1]
	{\frenchspacing
	 \ninerm\baselineskip=11pt
	 \begin{list}{\arabic{enumi}.}
	{\usecounter{enumi}\setlength{\parsep}{0pt}
	 \setlength{\leftmargin 12.7pt}{\rightmargin 0pt} %FOR 1--9 ITEMS
	 \setlength{\itemsep}{0pt} \settowidth
	{\labelwidth}{#1.}\sloppy}}{\end{list}}

\newcounter{itemlistc}
\newcounter{romanlistc}
\newcounter{alphlistc}
\newcounter{arabiclistc}

\newcommand{\fcaption}[1]{
        \refstepcounter{figure}
        \setbox\@tempboxa = \hbox{\footnotesize Fig.~\thefigure. #1}
        \ifdim \wd\@tempboxa > 5in
           {\begin{center}
        \parbox{5in}{\footnotesize\smalllineskip Fig.~\thefigure. #1}
            \end{center}}
        \else
             {\begin{center}
             {\footnotesize Fig.~\thefigure. #1}
              \end{center}}
        \fi}

\newcommand{\tcaption}[1]{
        \refstepcounter{table}
        \setbox\@tempboxa = \hbox{\footnotesize Table~\thetable. #1}
        \ifdim \wd\@tempboxa > 5in
           {\begin{center}
        \parbox{5in}{\footnotesize\smalllineskip Table~\thetable. #1}
            \end{center}}
        \else
             {\begin{center}
             {\footnotesize Table~\thetable. #1}
              \end{center}}
        \fi}

\def\@citex[#1]#2{\if@filesw\immediate\write\@auxout
	{\string\citation{#2}}\fi
\def\@citea{}\@cite{\@for\@citeb:=#2\do
	{\@citea\def\@citea{,}\@ifundefined
	{b@\@citeb}{{\bf ?}\@warning
	{Citation `\@citeb' on page \thepage \space undefined}}
	{\csname b@\@citeb\endcsname}}}{#1}}

\newif\if@cghi
\def\cite{\@cghitrue\@ifnextchar [{\@tempswatrue
	\@citex}{\@tempswafalse\@citex[]}}
\def\citelow{\@cghifalse\@ifnextchar [{\@tempswatrue
	\@citex}{\@tempswafalse\@citex[]}}
\def\@cite#1#2{{$\null^{#1}$\if@tempswa\typeout
	{IJCGA warning: optional citation argument 
	ignored: `#2'} \fi}}

\def\pmb#1{\setbox0=\hbox{#1}
	\kern-.025em\copy0\kern-\wd0
	\kern.05em\copy0\kern-\wd0
	\kern-.025em\raise.0433em\box0}

\def\fnt#1#2{\footnotetext{\kern-.3em
	{$^{\mbox{\scriptsize #1}}$}{#2}}}

\def\fpage#1{\begingroup
\voffset=.3in
\thispagestyle{empty}\begin{table}[b]\centerline{\footnotesize #1}
	\end{table}\endgroup}

\def\runninghead#1#2{\pagestyle{myheadings}
\markboth{{\protect\footnotesize\it{\quad #1}}\hfill}
{\hfill{\protect\footnotesize\it{#2\quad}}}}
\font\tenrm=cmr10
\font\tenit=cmti10 
\font\tenbf=cmbx10
\font\bfit=cmbxti10 at 10pt
\font\ninerm=cmr9

\font\eightrm=cmr8

\def\qed{\hbox{${\vcenter{\vbox{			%HOLLOW SQUARE
   \hrule height 0.4pt\hbox{\vrule width 0.4pt height 6pt
   \kern5pt\vrule width 0.4pt}\hrule height 0.4pt}}}$}}

\begin{document}

\runninghead{Discovery and Identification of $W'$ Bosons
$\ldots$} {Discovery and Identification of $W'$ Bosons
$\ldots$}

\normalsize\textlineskip
\thispagestyle{empty}
\setcounter{page}{1}

\copyrightheading{}			%{Vol. 0, No. 0 (1993) 000--000}

\vspace*{0.88truein}

\fpage{1}
\centerline{\bf DISCOVERY AND IDENTIFICATION OF $W'$ BOSONS}
\vspace*{0.035truein}
\centerline{\bf AT $e^+e^-$ COLLIDERS\footnote{This research was 
supported in part by the Natural Sciences and Engineering 
Research Council of Canada.   The 
work of M.A.D.\ was supported, in part, by the Commonwealth
College of The Pennsylvania State University under a 
Research Development Grant (RDG).}}
\vspace*{0.37truein}
\centerline{\footnotesize STEPHEN GODFREY, PAT KALYNIAK AND 
BASIM KAMAL}
\vspace*{0.015truein}
\centerline{\footnotesize\it 
Ottawa-Carleton Institute for Physics,}
\baselineskip=10pt
\centerline{\footnotesize\it 
Department of Physics, Carleton University, Ottawa Canada, K1S 5B6}
\vspace*{10pt}
\centerline{\footnotesize M.A. DONCHESKI}
\vspace*{0.015truein}
\centerline{\footnotesize\it 
Department of Physics, Pennsylvania State University}
\baselineskip=10pt
\centerline{\footnotesize\it Mont Alto, PA 17237 USA}
\vspace*{10pt}
\centerline{\footnotesize ARND LEIKE}
\vspace*{0.015truein}
\centerline{\footnotesize\it 
Ludwigs--Maximilians-Universit\"at, Sektion Physik, Theresienstr. 37,}
\baselineskip=10pt
\centerline{\footnotesize\it D-80333 M\"unchen, Germany}
%\vspace*{0.225truein}
%\publisher{(received date)}{(revised date)}

\vspace*{0.21truein}
\abstracts{We report on studies of the sensitivity to extra 
gauge bosons of the reactions $e^+ e^- \to \nu \bar{\nu} \gamma$ and
$e \gamma \to \nu q + X$ to extract discovery limits for $W'$'s.
The discovery potential for a $W'$ is, for some models, comparable to 
that of the LHC.  These processes may be also useful for 
determining $W'$ and $Z'$ couplings to fermions which would complement 
measurements made at the Large Hadron Collider.  }{}{}

\textlineskip			%) USE THIS MEASUREMENT WHEN THERE IS
\vspace*{12pt}			%) NO SECTION HEADING

\vspace*{1pt}\textlineskip	%) USE THIS MEASUREMENT WHEN THERE IS
%\section{Introduction}		%) A SECTION HEADING
%\vspace*{-0.5pt}
%\noindent
Extra  gauge bosons, both charged ($W'$) and/or neutral ($Z'$),
arise in many models of physics beyond the Standard 
Model (SM) \cite{c-g}. 
Examples  that we consider are the 
Left-Right symmetric model (LRM) based on the gauge group 
$SU(2)_L \times SU(2)_R \times U(1)_{B-L}$,
% which has right-handed charged currents, 
the Un-Unified model (UUM)
based on the gauge group $SU(2)_q \times SU(2)_l 
\times U(1)_Y$ where the quarks and leptons each transform under their own 
$SU(2)$, and the KK model (KK) which 
contains the Kaluza-Klein excitations of the SM gauge bosons that
are a possible consequence of theories with large extra dimensions. We 
also consider a $W'$ with SM couplings (SSM).
%The discovery of new gauge bosons would indicate that 
%the standard model gauge group was in need of extension.
In this contribution we give a brief summary of our work on
indirect searches for $W'$ bosons in $e^+e^-$ collisions. We are 
interested in two issues;  the sensitivity to $W'$ discovery and
the measurement of its couplings to thereby determine 
its origins. We refer 
the interested reader to the more detailed presentations of 
the process $e^+e^- \to \nu \bar{\nu} \gamma$ in Ref. 2 
and of the process $e \gamma \to \nu q +X$ in Ref. 3.

%\pagebreak

%\section{Calculations and Results}
%\noindent
The first process we consider is $e^+e^- \to \nu \bar{\nu} \gamma$. The 
kinematic variables of interest are the photon's energy, $E_\gamma$, 
and its angle relative to the incident electron, $\theta_\gamma$, both 
defined in the $e^+e^-$ centre-of-mass frame.  To take into account 
finite detector acceptance we imposed the constraints on the kinematic 
variables: $E_\gamma \geq 10$~GeV and $10^o \leq \theta_\gamma \leq 
170^o$.  The most serious background is radiative Bhabha scattering 
where the scattered $e^+$ and $e^-$ go undetected down the beam pipe.  
We suppress this background by restricting the photon's transverse 
momentum to $p_T^\gamma > \sqrt{s}\sin\theta_\gamma \sin\theta_v 
/(\sin\theta_\gamma +\sin\theta_v )$ where $\theta_v=25$~mrad and is 
the minimum angle to which the veto detectors may observe electrons or 
positrons.  There are also higher order backgrounds which cannot be 
suppressed, such as $e^+e^- \to \nu\bar{\nu} \nu' \bar{\nu}'\gamma$, 
and so must be included in an analysis of real data.

There is a large contribution to the cross-section from the radiative 
return of the $Z^0$ which is not sensitive to
$W'$'s and can be eliminated with a suitable cut on $E_\gamma$. 
%The low $E_\gamma$ region is most sensitive to 
%the presence of a $Z'$ because the low $E_\gamma$ region probes higher 
%$Z'$ masses in the $Z'$ propagator.  The cross-section corresponding 
%to right handed electrons exhibits a proportionality larger deviation 
%due to $Z'$ but because the right-handed cross-section is 
%significantly smaller than the left-handed cross section both left and 
%right handed cross-sections exhibit similar sensitivities.  On the 
%other hand, with any realistic degree of polarization the right-handed 
%cross-section will be swamped by the left-handed contributions.  
The statistical significance can be increased by binning the $E_\gamma$
distribution  and calculating the $\chi^2$. The limits 
obtained with and without a 2\% systematic error added 
in quadrature with the statistical error are given in Table 1.  

\begin{table}[t]
%\begin{table}[htbp]
\tcaption{$W'$ discovery limits in $TeV$. }
\centerline{\footnotesize\smalllineskip
\begin{tabular}{lllllllll}
\hline
&\multicolumn{4}{c}{$\sqrt{s}=0.5$ TeV, $L_{int}=500$ fb$^{-1}$} &
\multicolumn{4}{c}{$\sqrt{s}=1$ TeV, $L_{int}=500$ fb$^{-1}$}\\
&\multicolumn{2}{c}{$e^+e^-\rightarrow \nu\bar\nu\gamma$} 
&\multicolumn{2}{c}{$e\gamma\rightarrow \nu q + X$} &
 \multicolumn{2}{c}{$e^+e^-\rightarrow \nu\bar\nu\gamma$}
&\multicolumn{2}{c}{$e\gamma\rightarrow \nu q + X$}\\
Model &no syst.&syst.&no syst&syst.&no syst.&syst.&no syst.&syst.\\ \hline
SSM $W'$  & 4.3 & 1.7 & 4.1 & 2.6 & 5.3 & 2.2 & 5.8 & 4.2 \\
LRM       & 1.2 & 0.6 & 0.8 & 0.6 & 1.6 & 1.1 & 1.2 & 1.1 \\
UUM       & 2.1 & 0.6 & 4.1 & 2.6 & 2.5 & 1.1 & 5.8 & 4.2 \\
KK        & 4.6 & 1.8 & 5.7 & 3.6 & 5.8 & 2.2 & 8.3 & 6.0 \\ \hline
\end{tabular}}
\end{table}

The second process we consider is $e\gamma\to \nu q+X$ using
photon spectra from both the Weizsacker Williams process and from a 
backscattered laser.  Starting with the process $e\gamma \to 
\nu q\bar{q}$ the $W'$ contributions can be enhanced 
by imposing the kinematic cut that either the $q$ or $\bar{q}$ is 
collinear to the beam axis.  In this kinematic region the process 
$e\gamma\to \nu q\bar{q}$ is approximated quite well by the simpler 
process $e q\to \nu q'$ where the quark is described by the quark 
parton content of the photon, the so-called resolved photon 
approximation.  
%This has been verified numerically by comparing 
%kinematic distributions of the outgoing quark calculated using both 
%processes and with appropriate acceptance cuts on the observed and 
%unobserved $q(\bar{q})$.  
We use the process $eq\to \nu q'$ to obtain 
limits as it is computationally much faster and the limits obtained in 
this approximation are in good agreement with those using the full 
process.  The details of the calculation are given in Ref. 3.  

%\begin{figure}[htbp]
\begin{figure}[t]
\centerline{
\begin{minipage}[t]{6.0cm}
%\begin{figure}[htbp]
\vspace*{13pt}
\centerline{\hspace{-0.6cm}
\epsfig{file=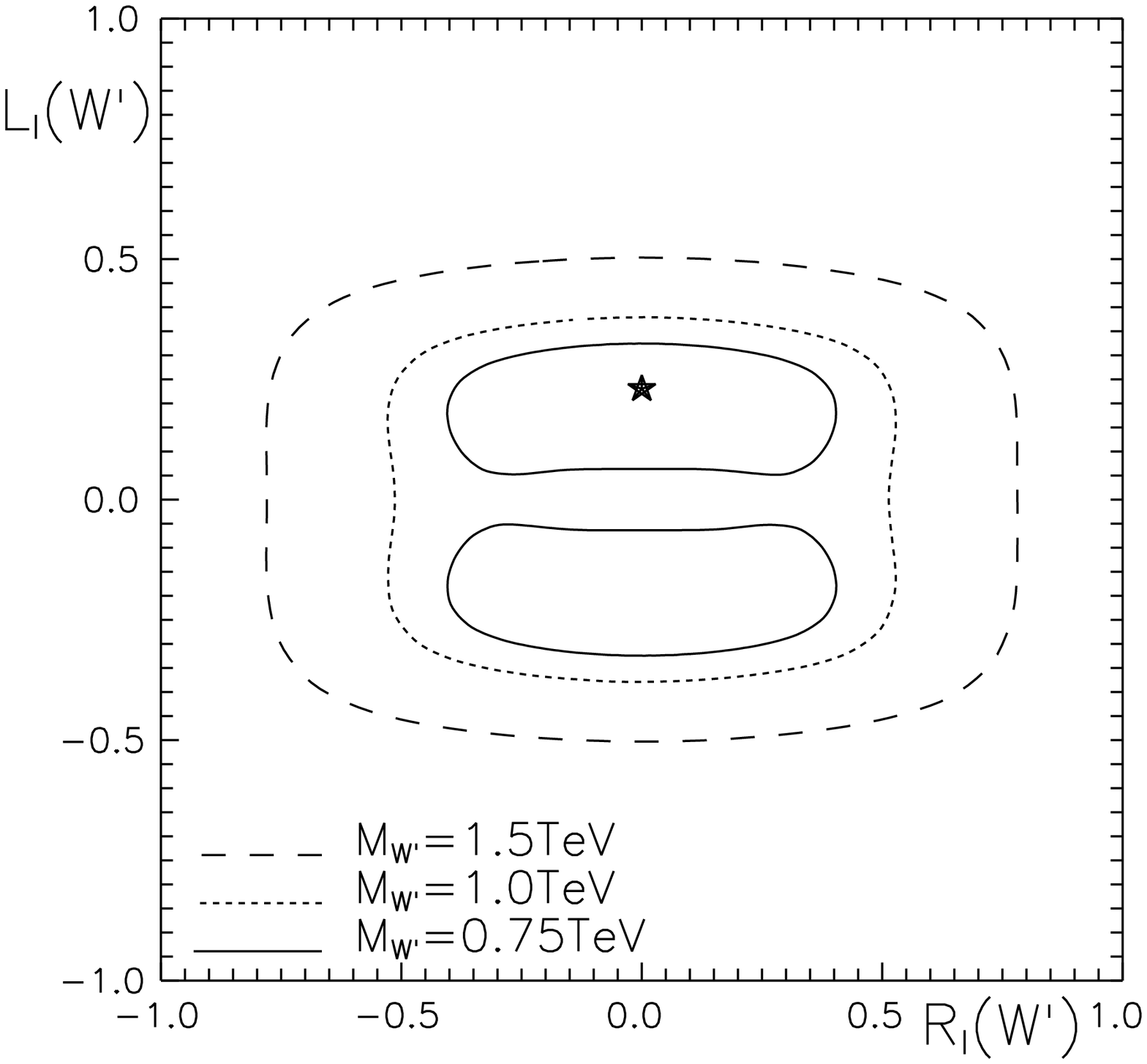,width=6.1cm,clip=}}
\centerline{(a)}
\vspace*{13pt}
%\end{figure}
\end{minipage} 
\hspace*{0.5cm}
\begin{minipage}[t]{6.0cm}
%\begin{figure}[htbp]
\vspace*{13pt}
\centerline{\hspace{-0.2cm}
\epsfig{file=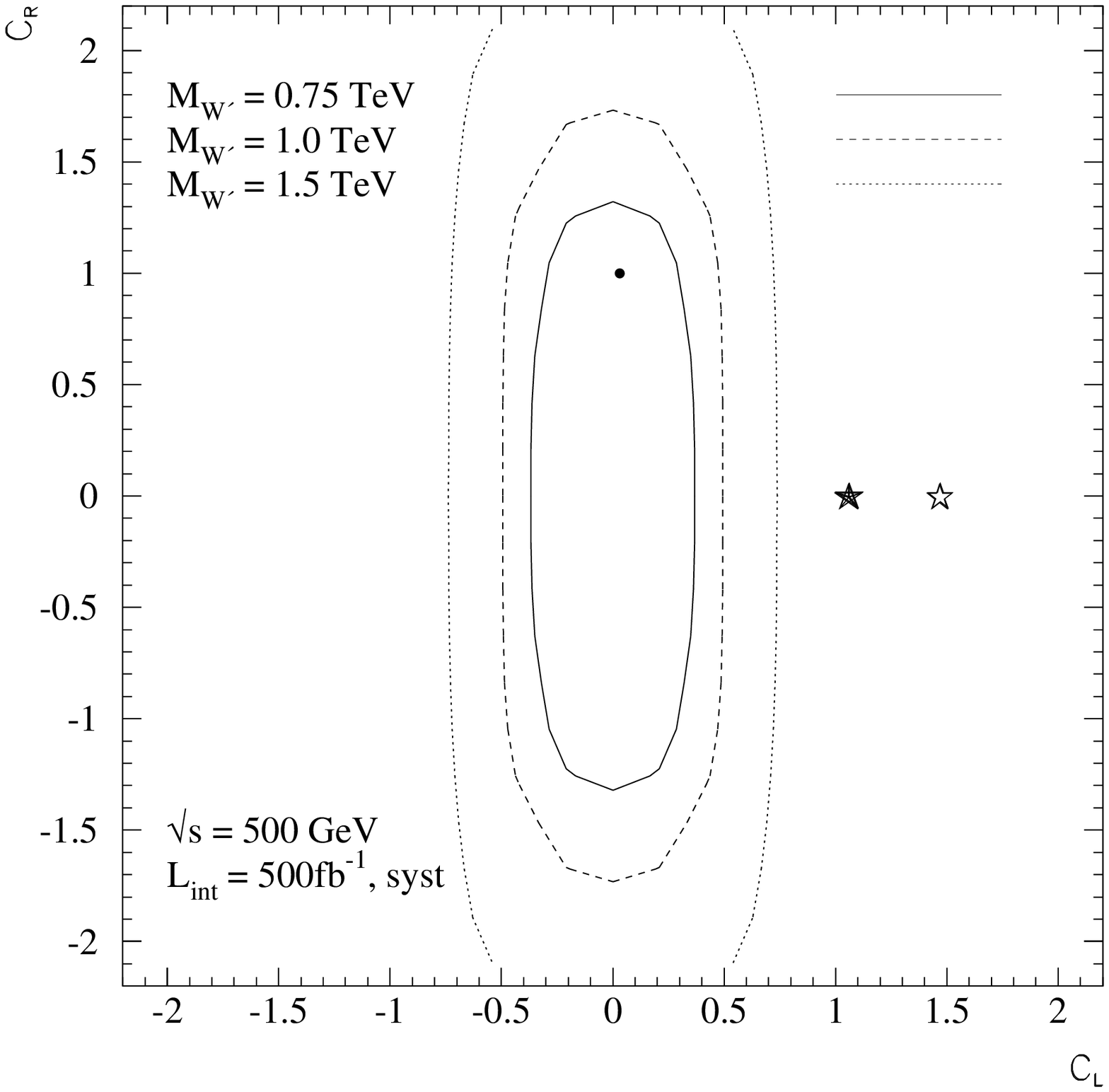,width=5.8cm,clip=}}
\centerline{(b)}
\vspace*{13pt}
%\end{figure}
\end{minipage}
}
\fcaption{
%\footnotesize{
%Fig. 1.  
95\% C.L. constraints on $W'$ couplings for $\sqrt{s}=0.5$~TeV 
and $L_{int}=500$~fb$^{-1}$ with systematic errors and
for different $W'$ masses.
(a) For the process $e^+e^-\to \nu\bar{\nu}\gamma$ assuming
a SSM $W'$ (indicated by a star)  using $\sigma$ and 
$A_{LR}$.  We take 90\% electron and 60\% positron polarization and 
include a systematic error of 2\% (1\% ) for $\sigma$ $(A_{LR})$.
(b) For the process $e\gamma \to \nu q +X$ 
with a backscattered laser spectrum  assuming the SM and 
using the $d\sigma/dp_{T_q}$ with a 2\% systematic 
error.  The SSM, LRM and the KK model are indicated 
by a full star, a dot and an open star, respectively.
(Note the different coupling normalizations in (a) and (b))
}
\end{figure}

To take into account detector acceptance
we restrict the angle of the outgoing $q(\bar{q})$ to the range $10^o \leq 
\theta_{q(\bar{q})} \leq 170^o$.  We have included $u$, $d$, $s$, and 
$c$-quark contributions and used the leading order GRV distributions.  
The search limits are fairly insensitive to the specific choice of 
distribution.  The kinematic variable most sensitive to $W'$
is the $p_{T_q}$ distribution.  The dominant backgrounds 
arise from two jet final states such as $\gamma \gamma \to q\bar{q}$,
$\gamma g \to q\bar{q}$, $gg \to q\bar{q}$ ..., 
where one of the jets goes down the beam pipe and is not observed. 
These backgrounds can be effectively eliminated by imposing the
constraint $p_{T_q} > 40 (75)$~GeV for $\sqrt{s}=0.5 (1.0)$~TeV. 
Discovery limits were obtained by binning the 
$p_{T_q}$ distribution and calculating the $\chi^2$ for an assumed 
integrated luminosity.  As before, a 2\%  systematic error was 
included in quadrature with the statistical error.  The discovery 
limits using the backscattered laser spectrum are given in Table 1.

If a signal was already detected for a $W'$ with mass lower 
than its search limit 
we can use the processes $e^+e^-\to \nu\bar{\nu}\gamma$
and $e\gamma \to \nu q +X$ to put constraints on the $W'$ couplings.  
%Constraints correspond to 
%95\% C.L. contours for agreement with the SM.  
In Fig. 1 we show constraints on $W'$ couplings for a collider with 
$\sqrt{s}=500$~GeV and $L_{int}=500$~fb$^{-1}$ for different $W'$ 
masses.
The constraints in Fig. 1a are for 
the process  $e^+e^-\to \nu\bar{\nu}\gamma$ 
found by combining the observables 
$\sigma$ and $A_{LR}$ for the process.  The 
axes correspond to couplings normalized as $L_f(W)=
C_L^{W_i}g/(2\sqrt{2})$ and similarly for $R_f(W)$.  Fig. 1b 
shows the constraints on $C_L$ and $C_R$ found by binning the $p_{T_q}$ 
distribution using the process 
$e\gamma \to \nu q +X$.  In Fig. 1b figure we have taken
$C_{L(R)}^e=C_{L(R)}^q$ which is satisfied in many models.

%\section{Summary}
%\noindent
We have shown that measurements at high energy $e^+e^-$ colliders 
are sensitive to $W'$ bosons much higher in mass than their 
centre-of-mass energy.  For some models the discovery reach is 
competitive with the LHC.  If a $W'$ were discovered, measurements of 
its couplings in the processes $e^+e^- \to \nu\bar{\nu}\gamma$ and 
$e\gamma\to \nu q +X$ would provide a valuable complement to 
measurements at the LHC.

\nonumsection{References}


\begin{thebibliography}{000}

\bibitem{c-g}
%For a recent reviews see
M.\ Cveti\u{c} and S.\ Godfrey; in {\it Electroweak Symmetry Breaking 
and Beyond the 
Standard Model}, eds.\ T.\ Barklow {\it et al.} (World Scientific, 
1995), p. 383, [hep-ph/9504216];
A.\ Leike, Phys.\ Rept.\ {\bf 317}, 143 (1999); %  See also
S.\ Godfrey, Phys.\ Rev.\ D {\bf 51}, 1402 (1995).

\bibitem{wp1}
S. Godfrey, P. Kalyniak, B. Kamal, and A. Leike,
Phys. Rev.  {\bf D61}, 113009 (2000) [hep-ph/0001074].

\bibitem{wp2}
S. Godfrey, P. Kalyniak, B. Kamal, M.A. Doncheski, and A. Leike,
[hep-ph/0008157].

\end{thebibliography}
\end{document}